\documentclass[prd,preprint,nofootinbib]{revtex4-1} 

\usepackage{latexsym,comment,verbatim}
\usepackage{amssymb}
\usepackage{amsfonts}
\usepackage{amsmath,color}
\usepackage[dvips]{graphicx}
\usepackage{bm}

\usepackage{natbib}

\usepackage{hyperref}

\hypersetup{
    colorlinks=true,
    linkcolor=red,
    citecolor=green,
    filecolor=magenta,      
    urlcolor=cyan,
    pdftitle={GWs},
    pdfpagemode=FullScreen,}

\def\mb#1{\mathbf{#1}}

\def\ber{\begin{eqnarray}}
\def\eer{\end{eqnarray}}
\def\beq{\begin{equation}}
\def\eeq{\end{equation}}

\def\rmd{{\rm d}}

\def\ed{\end{document}}

\def\dtau#1{\frac{\mathrm{d} #1}{\mathrm{d}\tau}}
\def\dttau#1{\frac{\mathrm{d} ^{2}#1}{\mathrm{d}\tau^{2}}}
\def\sT{\sin \left(\omega T \right)}
\def\cT{\cos \left(\omega T \right)}

\newcommand{\ppar}[2]{\frac{\partial #1}{\partial #2}}

\begin{document}

\author{Matteo Luca Ruggiero}
\email{matteoluca.ruggiero@unito.it}
\affiliation{Dipartimento di Matematica ``G.Peano'', Universit\`a degli studi di Torino, Via Carlo Alberto 10, 10123 Torino, Italy}
\affiliation{INFN - LNL , Viale dell'Universit\`a 2, 35020 Legnaro (PD), Italy}
\date{\today}

\title{Gravitomagnetic induction in the field of a gravitational wave}

\begin{abstract}
The interaction of a plane gravitational wave with test masses can be described in the proper detector frame, using Fermi coordinates, in terms of  a gravitoelectric and a gravitomagnetic field. We use this approach to calculate the displacements produced by  gravitational waves up to second order in the distance parameter and, in doing so, we emphasize the relevance of the gravitomagnetic contribution related to gravitational induction. In addition, we show how this approach can be generalized to calculate displacements up to arbitrary order.
\end{abstract}

\maketitle

\section{Introduction} \label{sec:intro}

The first direct detection \cite{abbott2016observation}  of gravitational waves (GW) took place 100 years after the birth of Einstein's theory of gravitation, and it was yet another success for General Relativity (GR). As a matter of fact, GR passed with flying colours many observational tests (see e.g. \citet{2014arXiv1409.7871W}) even though challenges come from  large scale observations of the Universe \cite{universe2040023} and we do not know yet how the reconcile it  with  the Standard Model  of particle physics. 

Nonetheless, the first observation of GW marked the beginning of gravitational wave astronomy and cosmology.  Besides being a test for the theory, today GW constitute a  tool to explore the Universe in the era of multi-messenger astronomy: technological developments and dedicated missions will help to greatly improve the information that can be obtained within this window (see e.g. \citet{2019Natur.568..469M,2021NatRP...3..344B}  and references therein). 

In this context, it is of utmost importance to define the measurement process: as it usually happens in GR, physical measurements are meaningful only when the observer and the object of the observations are unambiguously identified \cite{de2010classical}. The following steps summarize the measurement process in GR: (i) observers possess their own space-time, in the vicinity of their world-lines; (ii) covariant physics laws are  projected onto local space and time; (iii) predictions for the outcome of measurements  in the local space-time of the observers are obtained. When dealing with GW, there are two approaches to the description of the measurement process: the TT frame and the proper detector frame (see e.g. \citet{maggiore2007gravitational,Rakhmanov_2014}). 

The TT frame is based on the use  of a \textit{transverse and traceless} (TT)  tensor to describe the field of a gravitational wave, which allows to introduce the so-called TT coordinates; using these coordinates, the GW spacetime has a very simple form, and there are no gauge-depending information. On the other hand, these coordinates do not have a physical meaning: in fact the TT coordinates of a test mass in the field of a gravitational wave do not change or, differently speaking, particles which were at rest before the passage of the wave remain at rest after its  arrival (see e.g. \citet{Ruggiero:2021qnu}).  Of course, things are different if we consider the physical distances between test masses, which is modified by the passage of the wave, as it is testified by interferometric detection.  

The proper detector frame is based on the construction of Fermi coordinates, introduced in a seminal paper by Enrico Fermi \cite{1922RendL..31...21F}: these are a quasi-Cartesian coordinates system that can be build in the neighbourhood of a given observer, and their definition depends both on the background field where the observer is moving and, also, on his  motion. Fermi coordinates are  defined, by construction, as scalar invariants \cite{synge1960relativity}; they have a concrete meaning, since they are the coordinates an observer would naturally use to make space and time measurements in the vicinity of his world-line.

In previous works we showed that it is possible to describe the effects of a plane gravitational wave on the basis of a \textit{gravitoelectromagnetic} analogy \cite{Ruggiero_2020,Ruggiero:2021qnu}: differently speaking, the action of the wave on test masses can be explained in terms of a gravitoelectric and a gravitomagnetic field.  It is useful to remember that a gravitomagnetic field, produced by mass currents, naturally arises in GR both in the weak-field limit and in full theory (see e.g. \citet{Ruggiero:2002hz,Mashhoon:2003ax,Costa:2012cw}). Using this approach it is easy to understand that while current detectors reveal the interaction of test masses with the gravitoelectric components of the wave, there are also gravitomagnetic interactions that could be used to detect the effect of GW on moving masses  and spinning particles \cite{biniortolan2017,Ruggiero_2020b}.

The analogy that we used in previous works was limited to the description of displacements that are linear in the distance parameter; here, we further develop our approach to take into account quadratic displacements: we will show that these terms are related to \textit{gravitomagnetic induction} and they must be necessarily take into account. In doing so, we will rephrase the results obtained by \citet{baskaran} and discuss how to generalise them to arbitrary order.

The paper is organized as follows: we  review the gravitoelectromagnetic formalism applied to the field of a plane gravitational wave in Section \ref{sec:gem}, then we focus on its generalization to describe quadratic displacements in Section \ref{sec:correction}; discussion and conclusions are eventually in Section \ref{sec:disconc}.


\section{Gravitoelectric and gravitomagnetic fields arising from Fermi coordinates} \label{sec:gem}

Fermi coordinates in the vicinity of an observer's  world-line were studied in great details by \citet{Ni:1978di,Li:1979bz,1982NCimB..71...37F,marzlin}. The expression of the spacetime element in Fermi coordinates  depends both on the properties of the local reference frame (the world-line acceleration and the tetrad rotation)  and on the spacetime curvature, through the Riemann curvature tensor (see e.g. \citet{Ruggiero_2020}). Since we are interested in
gravitational waves effects, here  we  consider a geodesic and non rotating frame. Using  Fermi coordinates $(cT,X,Y,Z)$, up to quadratic displacements $|X^{i}|$ from the reference world-line, the line element turns out to be\footnote{Latin indices refer to space coordinates, while Greek indices to spacetime ones. Moreover,  we will use bold-face symbols like  $\mb W$ to refer to vectors in the Fermi frame.} (see e.g.  \citet{manasse1963fermi,MTW})
\beq
ds^{2}=-\left(1+R_{0i0j}X^iX^j \right)c^{2}dT^{2}-\frac 4 3 R_{0jik}X^jX^k cdT dX^{i}+\left(\delta_{ij}-\frac{1}{3}R_{ikjl}X^kX^l \right)dX^{i}dX^{j}. \label{eq:mmmetric}
\eeq
In the above equation  $R_{\alpha \beta \gamma \delta}(T)$ is the projection of the 
Riemann curvature tensor on the orthonormal tetrad $e^{\mu}_{(\alpha)}(\tau)$ of the
reference observer, parameterized by the proper time\footnote{In $e^{\mu}_{(\alpha)}$ tetrad indices like $(\alpha)$ are within parentheses, while  $\mu$ is a  background spacetime index; however, for the sake of simplicity, we drop here and henceforth parentheses to refer to tetrad indices, which are the only ones used.} $\tau$: $\displaystyle R_{\alpha \beta \gamma \delta}(T) = R_{\alpha \beta \gamma \delta}(\tau)=R_{\mu\nu \rho
\sigma}e^\mu_{(\alpha)}(\tau)e^\nu_{(\beta)}(\tau)e^\rho_{(\gamma)}(\tau)e^\sigma 
_{(\delta)}(\tau)$ and it is evaluated along the reference geodesic, where $T=\tau$ and $\mb X=0$.

By setting
\[
\frac{\Phi}{c^{2}}=\frac{g_{00}+1}{2} \quad \frac{\Psi_{ij}}{c^{2}}=\frac{g_{ij}-\delta_{ij}}{2} \quad \frac{A_{i}}{c^{2}}=-\frac{g_{0i}}{2},
\]
the above metric can be written in the form
\beq
\mathrm{d} s^2= -c^2 \left(1-2\frac{\Phi}{c^2}\right)\rmd T^2 -\frac4c A_{i}\rmd X^{i}\rmd T  +
 \left(\delta_{ij}+2\frac{\Psi_{ij}}{c^2}\right)\rmd X^i \rmd X^j\ , \label{eq:weakfieldmetric11}
\eeq
with the following defintions
\begin{eqnarray}
\Phi (T, { X^{i}})&=&-\frac{c^{2}}{2}R_{0i0j}(T )X^iX^j, \label{eq:defPhiG}\\
A^{}_{i}(T ,{X^{i}})&=&\frac{c^{2}}{3}R_{0jik}(T )X^jX^k, \label{eq:defAG}\\
\Psi_{ij} (T, {X^{i}}) & = & -\frac{c^{2}}{6}R_{ikjl}(T)X^{k}X^{l}, \label{eq:defPsiG}
\end{eqnarray}
where $\Phi$ and $A_{i}$ are, respectively, the \textit{gravitoelectric} and \textit{gravitomagnetic} potential, and $\Psi_{ij}$ is the perturbation of the spatial metric. Notice that the line element (\ref{eq:weakfieldmetric11}) is a perturbation of flat Minkowski spacetime; in other words $|\frac{\Phi}{c^{2}}| \ll 1$, $|\frac{\Psi_{ij}}{c^{2}}| \ll 1$, $|\frac{A_{i}}{c^{2}}| \ll 1$.  

In order to better understand the meaning of these potentials, we use them to write the geodesic equation. Namely, we start from the  line element (\ref{eq:weakfieldmetric11}) and calculate the geodesic equation
\beq
\displaystyle \dttau x^{\mu}+\Gamma^{\mu}_{\alpha\beta} \dtau{x^{\alpha}} \dtau{x^{\beta}}=0 \label{eq:geo00}
\eeq
up to linear order in ${\bm\beta}={\mathbf V}/c$, where $\displaystyle V^{i}=\frac{\rmd  X^i}{\rmd T}$. In analogy with electromagnetism, it is possible to define the \textit{gravitomagnetic field}
\beq
\mb B= \bm \nabla \wedge \mb A, \label{eq:defB}
\eeq
or, in terms of the curvature tensor
\beq
B^{}_i(T ,{\mb R})=-\frac{c^{2}}{2}\epsilon_{ijk}R^{jk}_{\;\;\;\; 0l}(T )X^l. \label{eq:defB000}
\eeq
Accordingly, the space components of the geodesic equation are
\beq
\frac{\rmd^{2} X^i}{\rmd T^{2}}=\frac{\partial \Phi}{\partial X^{i}}-2({\bm \beta}\times {\mathbf B})_i+2\frac{\partial A_{i}}{c\partial T}-2\beta^j \frac{\partial \Psi_{ij} }{c\partial T}-\beta^i \frac{\partial \Phi }{c\partial T}. \label{eq:geonew}
\eeq
In addition, exploiting once again the analogy with electromagnetism, we define the \textit{gravitoelectric field}
\beq
 \quad \mb E= -\bm \nabla \Phi-\frac{2}{c} \ppar{\mb A}{T}, \label{eq:defEtime}
\eeq
so that Eq. (\ref{eq:geonew}) becomes
\beq
\frac{\rmd^{2} X^i}{\rmd T^{2}}=-{ E}^{i}-2 \left(\frac{{\mathbf V}}{c}\times {\mathbf B}\right)^{i}-2\frac{V^{j}}{c} \frac{\partial \Psi_{ij} }{c\partial T}-\frac{V^{i}}{c} \frac{\partial \Phi }{c\partial T}.  \label{eq:lor2}
\eeq
Let us examine the meaning of  Eq. (\ref{eq:lor2}) and  relevance of the various terms.  First of all, it is important to stress that this equation defines the motion of a test mass with respect to the reference observer. Consequently, all quantities involved are \textit{relative} to the reference observer at the origin of the frame. 
In addition, the geodesic equation does not take a Lorentz-like form if the fields are not static, due to the presence of the last terms which contain time-derivatives \cite{Ruggiero:2021uag}. However,  both terms - according to the definitions (\ref{eq:defPhiG}) and (\ref{eq:defPsiG}) - are quadratic in the displacements from the reference world-line. Hence, up to linear displacements, we obtain the Lorentz-like force
\beq
\frac{\rmd^{2} \mb X}{\rmd T^{2}}=-{ \mb E}-2 \left(\frac{{\mathbf V}}{c}\times {\mathbf B}\right),  \label{eq:lor2lin}
\eeq
where the gravitoelectric field  in this case  is
\beq
\mb E= -\bm \nabla \Phi. \label{eq:defEzero}
\eeq

There is another interesting regime for Eq. (\ref{eq:lor2}), which is relevant for gravitational waves: this is the case when the test masses are at rest before the passage of the wave. In fact, if we work at linear order in the wave amplitude,  we can neglect  terms proportional to $\mb V/c$, and the force equation is
\beq
\frac{\rmd^{2} \mb X}{\rmd T^{2}}=-{ \mb E}.  \label{eq:lor2lin2}
\eeq
We may solve this equation using the definition of the gravitoelectric field (\ref{eq:defEtime}). However, in doing so, care must be paid to the definition of the gravitoelectric potential $\Phi$: in fact, if we use the definition (\ref{eq:defPhiG}), we obtain
\beq
E^{}_i(T ,{\mb X})=c^{2}R_{0i0j}(T) X^j-\frac 2 3 c \frac{\partial R_{0jik}(T)}{\partial T} X^{j}X^{k}. \label{eq:defE0}
\eeq
In this equation, we see that while the contribution coming from $\Phi$ is \textit{linear} in the displacements from the reference world-line, the gravitomagnetic contribution is \textit{quadratic.}  Accordingly, we need to develop $g_{00}$ in (\ref{eq:mmmetric}) up to \textit{cubic} displacements from the reference world-line. The result turns out to be (see e.g. \citet{Li:1979bz,1982NCimB..71...37F,marzlin,Rakhmanov_2014})

\beq
\Phi=-\frac{c^{2}}{2}  R_{0i0j}(T) X^{i}X^j-\frac{c^{2}}{6}R_{0i0j,m}(T) X^{i}X^jX^{m}, \label{eq:Phirev}
\eeq
where $\displaystyle R_{0i0j,m} (T) = \frac{\partial R_{0i0j}}{\partial X^{m}}$, and this expression is evaluated along the reference geodesic, where $T=\tau$ and $\mb X=0$. For future convenience, we will use these definitions:
\beq
\Phi=\Phi^{(0)}+\Phi^{(1)}\quad \Phi^{(0)}=\frac{c^{2}}{2}  R_{0i0j}(T) X^{i}X^j, \quad \Phi^{(1)}=-\frac{c^{2}}{6}R_{0i0j,m}(T) X^{i}X^jX^{m}. \label{eq:Phisdef}
\eeq

Below, we will focus on the impact of the gravitomagnetic contribution in (\ref{eq:defEtime}), which can be seen as a \textit{induction term.} This is easy to understand, if we remark the meaning of the gravitoelectromagnetic analogy, which is summarised by  equation (\ref{eq:lor2lin}):  this equation  describes the evolution of a test mass, i.e. how its spatial coordinates $X,Y,Z$ change according to the reference observer. Then, the action of a gravitational field is simply described in terms of  forces: the latter are due to the presence of gravitoelectromagnetic fields which, because of their definitions (\ref{eq:defB}) and (\ref{eq:defEtime}) satisfy the homogeneous Maxwell equations
 \begin{eqnarray}
 \bm \nabla \times \mb E+\frac 2 c \ppar{\mb B}{T} &=&0, \label{eq:constE} \\
 \bm \nabla \cdot \bm B &=& 0. \label{eq:constB}
 \end{eqnarray}
On the other hands, it is easy to show that these fields {\textit{satisfy the inhomogeneous Maxwell equations with source terms, even though they are vacuum solutions}} (see e.g. \citet{Mashhoon:2003ax}), which evidently breaks the analogy, whose ultimate meaning  is the possibility to write the geodesic equation in the form (\ref{eq:lor2lin}) and to obtain  consequences similar to those that we know in electromagnetism.   In particular, we see that the induction law (\ref{eq:constE}) is a direct consequence of the definition of the gravitoelectric field (\ref{eq:defEtime}).

\subsection{Gravitoelectromagnetic formalism for a plane gravitational wave} \label{ssec:GMGW}

Here, we apply the formalism described above to the field of a plane gravitational wave; in particular, we consider the limit defined by Eq. (\ref{eq:lor2lin}), i.e. up to linear displacements from the reference geodesic. The metric can be written in the form $g_{\mu\nu}=\eta_{\mu\nu}+h_{\mu\nu}$, where $h_{\mu\nu}$ is a perturbation of the flat Minkowski spacetime $\eta_{\mu\nu}$. Up to linear order in the perturbation $h_{\mu\nu}$, we can write the following expressions for the Riemann tensor \cite{MTW}:
\beq
R_{ikjl}=\frac 1 2 \left(h_{il,jk}+h_{kj,li}-h_{kl,ji}-h_{ij,lk} \right) \label{eq:riemann0}
\eeq
and
\beq
R_{ij0l}=\frac 1 2 \left( h_{il,j0}-h_{jl,i0} \right). \label{eq:riemann1}
\eeq
In particular, we consider the TT coordinates $(ct,x,y,z)$, so that the components for a wave propagating along the $x$ direction are \cite{Ruggiero:2021qnu}
\beq
h_{xx}=1, \quad h_{yy}=1-A^{+}\sin\left(\omega t-kx\right), \quad  h_{zz}=1+A^{+}\sin\left(\omega t-kx\right), \quad  h_{zy}=-A^{\times}\cos\left(\omega t-kx\right), \label{eq:perturb}
\eeq
where  $\omega$ is the frequency and $k$ the wave number, so that the wave four-vector is $\displaystyle k^{\mu}=\left(\frac \omega c, k, 0, 0 \right)$, with $k^{\mu}k_{\mu}=0$; $A^{+}, A^{\times}$ are the amplitudes of the wave in the two polarization states. We exploit the gauge invariance in linear approximation \cite{straumann2013applications}  and use the above expressions to calculate the Riemann tensor in Fermi coordinates; notice that in the metric (\ref{eq:mmmetric}) the Riemann tensor is evaluated along the reference world-line: accordingly, after calculating the components of Riemann tensor using Eq. (\ref{eq:perturb}), we set $X=0$.

The gravitoelectric potential (\ref{eq:defPhiG}) is
\beq
\Phi=\frac{\omega^{2}}{4}\left[ A^{+}\sT Y^{2}+2A^{\times}\cT YZ-A^{+}\sT Z^{2} \right], \label{eq:defPhicomp}
\eeq
while the components of the gravitomagnetic potential (\ref{eq:defAG})  are
\begin{eqnarray}
A_{X}&=&\frac{\omega^{2}}{6} \left[A^{+}\sT (Y^{2}-Z^{2})+2A^{\times}\cT ZY \right], \label{eq:defAX} \\
A_{Y}&=& \frac{\omega^{2}}{6} \left[-A^{+}\sT YX-A^{\times}\cT ZX \right], \label{eq:defAY} \\
A_{Z}&=& \frac{\omega^{2}}{6} \left[-A^{\times}\cT YX+A^{+}\sT XZ \right].  \label{eq:defAZ}
\end{eqnarray}
From the above expressions, we obtain the components of the gravitoelectric field (\ref{eq:defEzero}): 
\small
\beq
E^{}_{X}  = 0, \quad E^{}_{Y}  = -\frac{\omega^{2}}{2}\left[A^{+} \sin \left(\omega T \right)Y+A^{\times} \cos \left(\omega T \right) Z \right], \quad E^{}_{Z}  = -\frac{\omega^{2}}{2}\left[A^{\times}\cT Y-A^{+}\sT Z \right], \label{eq:campoE}
\eeq
\normalsize
and those of the gravitomagnetic field  (\ref{eq:defB}):
\small
\beq
B^{}_{X}  = 0, \quad B^{}_{Y}  = -\frac{\omega^{2}}{2}\left[-A^{\times} \cT Y+A^{+} \sT Z \right], \quad B^{}_{Z}  = -\frac{\omega^{2}}{2}\left[A^{+}\sT Y+A^{\times}\cT Z \right]. \label{eq:campoB}
\eeq
\normalsize
%
%

We see that, up to linear displacements, both fields are transverse to the propagation direction. The effect on test masses at rest is then determined by the gravitoelectric field (\ref{eq:campoE}): as a consequence,  there are no displacements along the propagation direction.

\section{The gravitomagnetic contribution to the gravitoelectric force} \label{sec:correction}

In this Section we generalise the approach described before to displacements that are quadratic in the distance parameter. Starting from the definition of the gravitoelectric field (\ref{eq:defEtime}), we may write
\beq
\mb E= \mb E^{(0)}+\mb E^{(1)} \label{eq:defE1}
\eeq
where $\displaystyle \mb E^{(0)}=-\bm \nabla \Phi^{(0)}$ derives from the quadratic part of the gravitoelectric potential (\ref{eq:Phisdef}) and, hence, is \textit{linear} in the displacement $|X^{i}|$ from the reference world-line. In particular, its expression was calculated in Section \ref{ssec:GMGW} in Eq. (\ref{eq:campoE}). In the  detector frame this gravitoelectric force can be thought of as ``Newtonian'', since it does not depend on the speed of light.

On the other hand, $\displaystyle \mb E^{(1)}=-\bm \nabla \Phi^{(1)}-\frac{2}{c} \ppar{\mb A}{T}$: it  derives from the cubic part of the gravitoelectric potential (\ref{eq:Phisdef})  and  from the gravitomagnetic induction term so, taking into account  Eq. (\ref{eq:defAG}), it is  \textit{quadratic} in the displacements $|X^{i}|$ from the reference geodesic. As we are going to show, this term is $O(1/c)$ and, in the detector frame, it can be thought of as the first relativistic correction to the dominant ``Newtonian'' force.
We obtain the following expression for $\Phi^{(1)}$
\beq
\Phi^{(1)}=-\frac{1}{12}\frac{\omega^{3}}{c}\left[A^{+}\cT \left(Y^{2}-Z^{2} \right)-2A^{\times}\sT YZ \right]X, \label{eq:defPhi1}
\eeq
which, taking into account Eq. (\ref{eq:defAX})-(\ref{eq:defAZ}), leads to the following expressions for $\mb E^{(1)}$:
\begin{eqnarray}
E^{(1)}_{X}&=&-\frac 1 4 \frac{\omega^{3}}{c}\left[A^{+}\cT \left(Y^{2}-Z^{2} \right)-2A^{\times}\sT Z Y \right], \label{eq:E1x} \\
E^{(1)}_{Y}&=&\frac 1 2\frac{\omega^{3}}{c}\left[A^{+}\cT YX-A^{\times}\sT ZX \right], \label{eq:E1y} \\
E^{(1)}_{Z}&=&-\frac 1 2 \frac{\omega^{3}}{c}\left[A^{\times}\sT YX+A^{+}\cT ZX \right]. \label{eq:E1z}
\end{eqnarray}

In particular, from (\ref{eq:E1x}) we see that a non null $X$ component of the gravitoelectric field is present, contrary to what happens at linear order (see   Eq. (\ref{eq:campoE})).  If $L$ denotes a typical detector distance and $\lambda$ is the wavelength, we see that
\beq
\frac{|\mb E^{(1)}|}{|\mb E^{(0)}|} \simeq \frac{\omega L}{c} \simeq \frac{L}{\lambda}. \label{eq:confronto}
\eeq
Accordingly, $\mb E^{(1)}$ can be thought of as the first relativistic correction, and it is smaller by a factor $\displaystyle \frac{L}{\lambda}$ with respect to the leading ``Newtonian'' term. 

Given the above expressions for $\mb E^{(1)}$, and taking into account the expression (\ref{eq:campoE}) for $\mb E^{(0)}$, we can  solve the equation
\beq
\frac{\rmd^{2} \mb X}{\rmd T^{2}}=-\mb E \label{eq:geobase}
\eeq
for a test mass with the initial conditions $\left(X_{0},Y_{0},Z_{0}\right)$ at $T=0$. We obtain the following  solution
\begin{eqnarray}
X(T)&=&X_{0}+\frac 1 4 \frac{\omega^{}}{c}A^{+}\left[1-\cT \right] \left(Y^{2}_{0}-Z^{2}_{0}\right)+\frac 1 2 \frac{\omega}{c}A^{\times}\sT Z_{0}Y_{0}, \label{eq:solXT}\\
Y(T) & = & \left[1-\frac{A^{+}}{2}\sT \right]Y_{0}+\frac{A^{\times}}{2}\left[1-\cT \right]Z_{0}+\frac{1}{2}\frac{\omega}{c}A^{+}\left[\cT-1\right]X_{0}Y_{0}+ \nonumber \\ &-&\frac 1 2 \frac{\omega}{c}A^{\times}\sT X_{0}Z_{0}, \label{eq:solYT} \\
Z(T) & = & \left[1+\frac{A^{+}}{2}\sT \right]Z_{0}+\frac{A^{\times}}{2}\left[1-\cT \right]Y_{0}+\frac 1 2 \frac{\omega}{c}A^{+}\left[1-\cT \right]X_{0}Z_{0}+ \nonumber \\ &-&\frac 1 2 \frac{\omega}{c}A^{\times}\sT X_{0}Y_{0}. \label{eq:solZT}
\end{eqnarray}

\begin{figure}[t]
\begin{center}
\includegraphics[scale=.2]{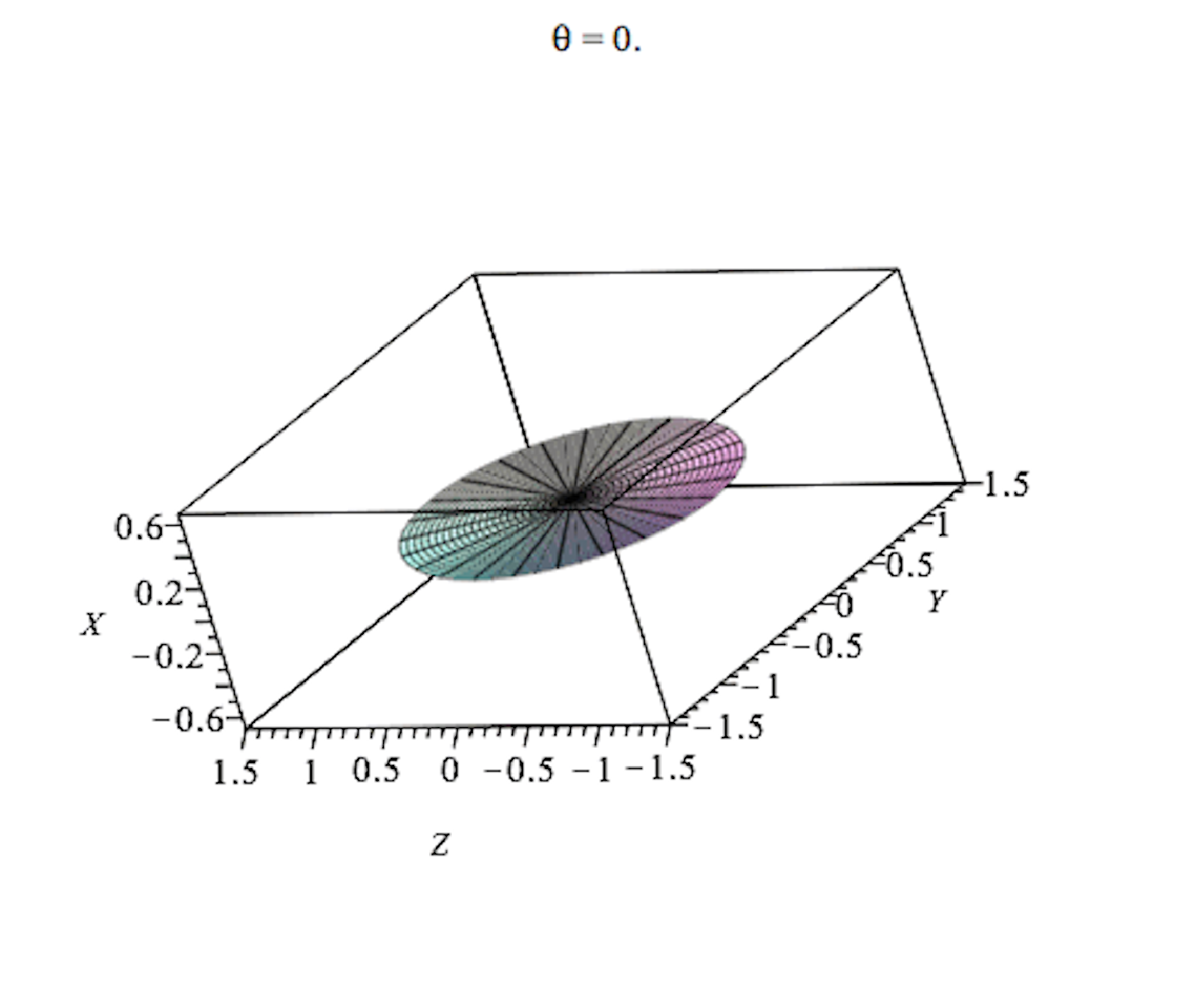}
\includegraphics[scale=.2]{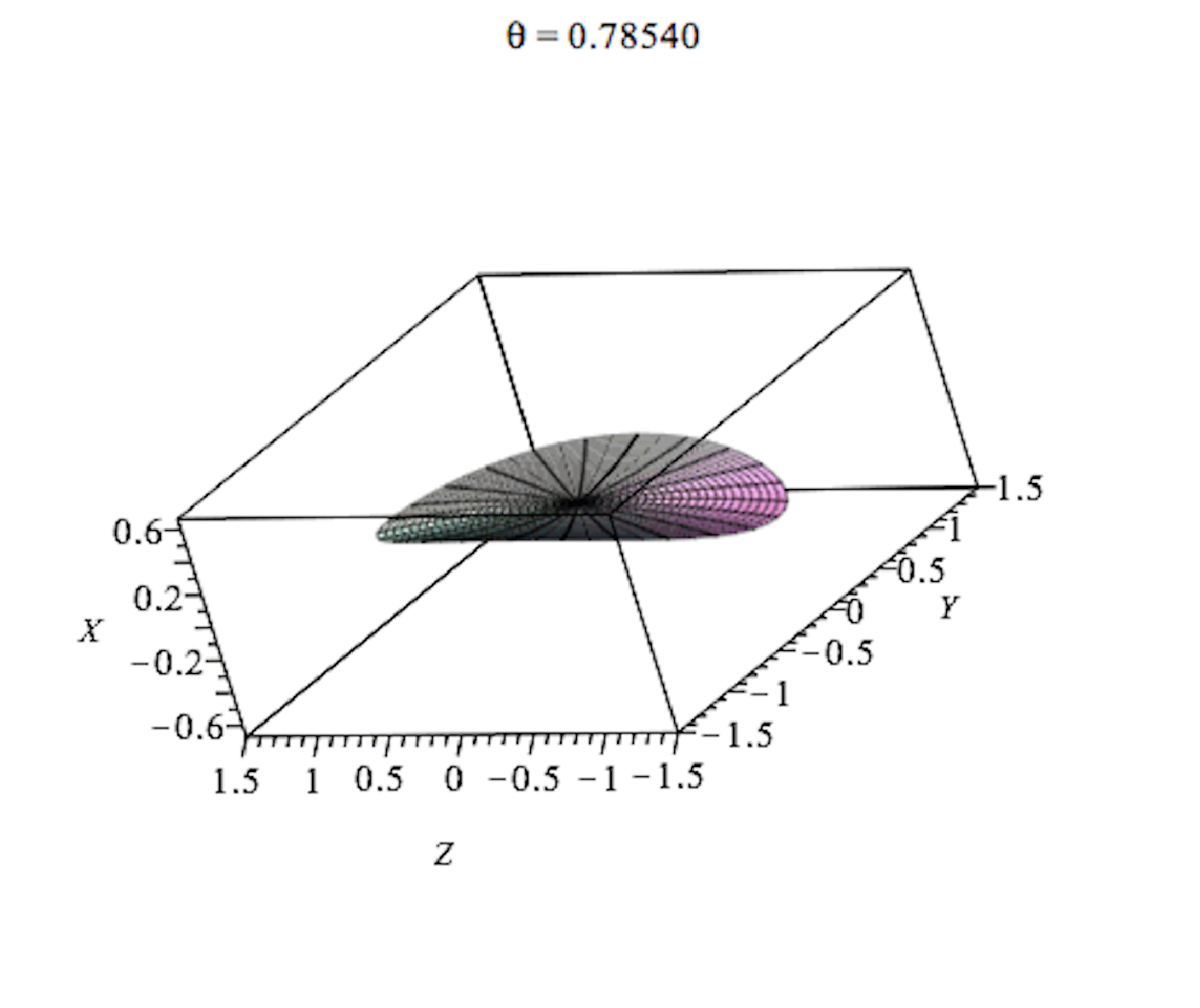}
\includegraphics[scale=.2]{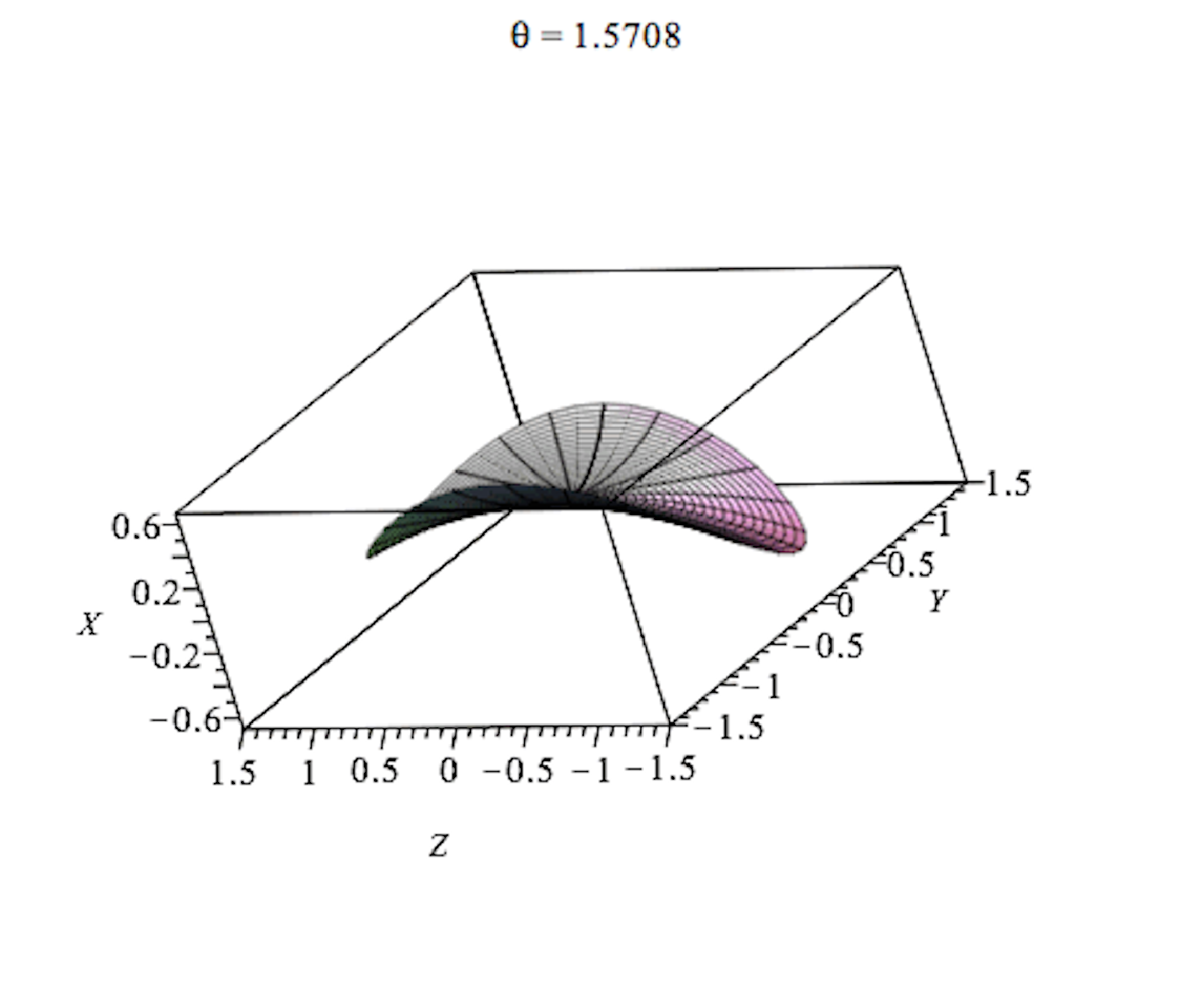}
\caption{Time evolution due to the gravitomagnetic effects of the $+$ polarization; $\theta=\omega T$. } \label{fig:figurepiu}
\end{center}
\end{figure}

Notice that if we neglect the contribution from $\mb E^{(1)}$, there is no effect along the propagation direction, i.e. $\displaystyle X(T)=X_{0}+O(\frac L \lambda)$.

Using the above expressions (\ref{eq:solXT})-(\ref{eq:solZT}), it is possible to visualise the deformation produced by the GW on a  disc of test masses, lying at $T=0$ on the $X=0$ plane: the results are depicted in Figure \ref{fig:figurepiu} for the $A^{+}$ polarization. In particular, due to the gravitomagnetic term, the disc gets deformed \textit{out of its plane}, contrary to what happens for linear displacements.

Starting from the above results, following \citet{baskaran}, we may calculate the distance between a test mass at the origin  and one lying at an arbitrary location $(X_{0},Y_{0},Z_{0})$ before the passage of the wave: this is a simple model for the displacements of two mirrors in an interferometer. For the second mass we may write:
\beq
X(T)=X_{0}+\delta X(T), \quad Y(T)=Y_{0}+\delta Y(T), \quad Z(T)=Z_{0}+\delta Z(T), \label{eq:deltas}
\eeq
where $\delta X(T), \delta Y(T), \delta Z(T)$ can be read from Eqs. (\ref{eq:solXT})-(\ref{eq:solZT}). Our analysis is limited to linear order in the wave amplitude: consequently,  setting $\displaystyle D(T)=\sqrt{X(T)^{2}+Y(T)^{2}+Z(T)^{2}}$, we obtain
\beq
D(T)=D_{0}+\frac{\delta X(T) X_{0}+\delta Y(T) Y_{0}+\delta Z(T) Z_{0}}{D_{0}} \label{eq:DT1}
\eeq
where $\displaystyle D_{0}=\sqrt{X_{0}^{2}+Y_{0}^{2}+Z_{0}^{2}}$ is the initial distance. 

Let us consider the case where the two test masses are in the plane orthogonal to the propagation direction: without loss of generality, we may consider the second test mass at $(0,Y_{0},0)$, whence $D_{0}=Y_{0}$. From (\ref{eq:DT1}) we have
\beq
D(T)=Y_{0}+\delta Y. \label{eq:DT2}
\eeq
As a consequence, we see that there is no impact of $\mb E^{(1)}$ and, hence, of the gravitomagnetic contribution. 

Things are different if the two test masses are out of  the plane perpendicular to the propagation direction: from (\ref{eq:DT1}) and (\ref{eq:solXT})-(\ref{eq:solZT}) we obtain the general expression for $D(T)$:
\begin{align}
D(t)&=D_{0}+ \frac{2A^{\times}Z_{0}Y_{0} \left(1-\cT \right)+A^{+}\left(Z_{0}^{2}-Y_{0}^{2}\right)\sT}{2D_{0}} \nonumber \\
 & +\frac{\omega X_{0}}{c}\frac{A^{+}\left(1-\cT \right)\left(Z_{0}^{2}-Y_{0}^{2} \right)-2A^{\times}\sT Z_{0}Y_{0}}{4D_{0}}. \label{eq:distgen}
\end{align}
The term in the second line represents a correction to the leading expression of the distance, and this term is smaller by $\displaystyle \frac{\omega X_{0}}{c} \simeq \frac{L}{\lambda}$: this correction term is present only when $X_{0} \neq 0$, so the masses should not be in the plane orthogonal to the propagation direction. The results are in agreement with \citet{baskaran}.

The approach described so far can be generalised to calculate the effects of GW up to arbitrary order in the distance parameter. To do this, it is sufficient to write the gravitoelectromagnetic potentials to all order in the distance parameter:  we may write (see \citet{Li:1979bz,1982NCimB..71...37F,marzlin,Rakhmanov_2014})
\beq
\Phi=-c^{2}\sum_{n=0}^{\infty} \frac{n+3}{(n+3)!} R_{0i0j,m_{1}...m_{n}}X^{i}X^{j}X^{m_{1}}...X^{m_{n}}, \label{eq:PhigenN}
\eeq
where $\displaystyle R_{ikjl,m_{1}...m_{n}}=\frac{\partial^{n} R_{ikjl}}{\partial X^{m_{1}}...\partial X^{m_{n}}}$ and this expression is evaluated along the reference geodesic, where $T=\tau$ and $\mb X=0$.

From (\ref{eq:PhigenN}) we may write $\Phi=\Phi^{(0)}+\Phi^{(1)}+...+\Phi^{(n)}+...$, where the $n$-th term is
\beq
\Phi^{(n)}=-c^{2}\frac{n+3}{(n+3)!} R_{0i0j,m_{1}...m_{n}}X^{i}X^{j}X^{m_{1}}...X^{m_{n}}. \label{eq:PhigenNn}
\eeq

As for the gravitomagnetic potential, we have
\beq
A_{i}=c^{2}\sum_{n=0}^{\infty} \frac{n+2}{(n+3)!} R_{0lij,m_{1}...m_{n}}X^{l}X^{j}X^{m_{1}}...X^{m_{n}}. \label{eq:AgenN}
\eeq
We may write $\mb A= \mb A^{(0)}+\mb A^{(1)}+...+\mb A^{(n)}+...$ where the $n$-th term is
\beq
\mb A^{(n)}=c^{2}\frac{n+2}{(n+3)!} R_{0lij,m_{1}...m_{n}}X^{l}X^{j}X^{m_{1}}...X^{m_{n}}.  \label{eq:AgenNn}
\eeq

Accordingly, the $n$-order correction to the gravitoelectric field can be written in the form
\beq
\mb E^{(n)}=-\bm \nabla \Phi^{(n)}-\frac 2 c \frac{\partial \mb A^{(n-1)}}{\partial T}, \label{eq:defEn}
\eeq
and it determines the correction at the $n+1$ order in the distance parameter. 

We remark that the expressions for the gravitoelectromagnetic potentials  (\ref{eq:PhigenN}) and (\ref{eq:AgenN}) to be used in the metric (\ref{eq:weakfieldmetric11}) can be written in a very compact form in the case of a plane gravitational wave \cite{1982NCimB..71...37F,Berlin:2021txa}.

\subsection{Gravitomagnetic Induction} \label{ssec:induction}

\begin{figure}[h]
\begin{center}
\includegraphics[scale=.7]{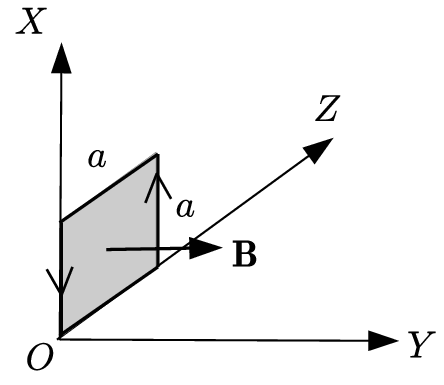}
\caption{The closed circuit is a square with side $a$; the gravitomagnetic field is always orthogonal to the square.} \label{fig:figura}
\end{center}
\end{figure}

As we have seen before, the displacements produced by GW  on test masses at  second order in the distance parameter are determined both by the gravitoelectric potential $\Phi_{1}$  and by the gravitomagnetic potential $\mb A$. The magnitude  of these effects is $\displaystyle \frac{L}{\lambda}$ smaller with respect to those at linear order. Our calculations show that  to consistently calculate these effects, gravitomagnetic terms cannot be neglected: in particular, according to Eq. (\ref{eq:defEtime}), a test mass $m$ initially at rest undergoes a gravitomagnetic force $\displaystyle \frac{2m}{c} \frac{\partial {\mathbf A}}{\partial T} $.

These effects can be thought of as a consequence of a gravitomagnetic induction (see e.g. \citet{bini2008gravitational}): in fact,  we have shown that the law of gravitational induction for gravitoelectromagnetic fields (\ref{eq:constE}) holds true. Actually, the latter equation can be written in integral form
\beq
\oint \mb E \cdot  \rmd {\pmb \ell}=-\frac{2}{c}\frac{\partial }{\partial T^{}} \int {\mathbf B}\cdot \rmd {\mathbf S}. \label{eq:indint}
\eeq
The l.h.s of the above equation, in analogy with electromagnetism where it represents the electromotive force (emf), is the work done per unit test mass. By its very definition, it is evident that only $\mb E^{(1)}$ contributes to Eq. (\ref{eq:indint}). 

We may evaluate the impact of this \textit{gravitomotive} force using a simple model. Let us consider as closed circuit the boundary of a square of side $a$ in the $Y=0$ plane (see Figure \ref{fig:figura}); we consider a  GW with $A^{\times}=0$.  The GM field (\ref{eq:campoB}) is in the form
\beq
B^{}_{X}  = 0, \quad B^{}_{Y}  = -\frac{\omega^{2}}{2}\left[A^{+} \sT Z \right], \quad B^{}_{Z}  = 0. \label{eq:campoB01}
\eeq
We obtain
\beq
\oint \mb E \cdot  \rmd {\pmb \ell}=\frac{\omega^{3}}{2c}A^{+}\sT a^{3}. \label{eq:gemforce}
\eeq
Accordingly, we may say that the displacements along the propagation direction can be explained in terms of gravitational induction: the gravitomagnetic field (\ref{eq:campoB}), which is orthogonal to the propagation direction, induces a gravitoelectric field parallel to the propagation direction.

\section{Discussion and conclusions} \label{sec:disconc}

We investigated the effects of plane gravitational waves on test masses using a gravitoelectromagnetic analogy that naturally emerges in the proper detector frame, using Fermi coordinates. Elsewhere \cite{Ruggiero_2020,Ruggiero_2020b} we discussed the effects of the magnetic-like part of the GW field on moving and spinning particles. Here we focused on the effects on test masses at rest before the passage of the wave, and showed that gravitomagnetic terms are relevant to describe the  displacements  up to second order in the distance parameter. In particular, while at first order the interaction of a detector with GW can be  explained in terms of the action of a gravitoelectric and gravitomagnetic field transverse to the propagation direction, at second order the gravitoelectric field has a non null component along the propagation direction. In addition, to calculate the gravitoelectric field  at second order, it is important to consider a contribution deriving from the gravitomagnetic potential, which is directly related to what we called gravitomagnetic induction. These corrections are smaller by a factor 
\[
\displaystyle \frac{|\mb E^{(1)}|}{|\mb E^{(0)}|} \simeq \frac{\omega L}{c} \simeq 8 \times 10^{-2} \left(\frac{L}{\mathrm{4\, km}} \right) \left(\frac{\nu}{\mathrm{1000\, Hz}} \right)
\]  
with respect to the first order terms, that are measured in current interferometers. We see that neglecting these term will have an impact of the same order of magnitude on the knowledge of the GW sources parameters,  which is especially relevant in the high-frequency regime.  In addition, we showed how our formalism can be generalized to calculate the displacements to arbitrary order in the distance parameter.  

Actually, it is important to point out that the response of an interferometer depends on two different contributions: besides the geometric displacements of  the two mirrors, there is the photons time delay \cite{fortiniortolan91,fortiniortolan92}; the latter contribution is generally negligible with respect to  space displacements, however it should be relevant for an analysis at second order in the distance parameter such as the one discussed in this paper. A comprehensive study of the interferometric response using the gravitoelectromagnetic formalism  will be object of future works.

Besides providing a tool for accurate evaluation of the effect of GW, our results could be relevant also to test theories of gravity alternative to GR. In fact, in these theories there are longitudinal effects in gravitational radiation due, for instance, to massive modes, scalar fields or to a richer geometric structure (see e.g. \citet{capozziello1,capozziello2,corda1,corda2}). The possible detection of these effects will be an important test for these theories: but this could be possible only if existing GR effects are accurately modelled and taken into account. {In particular, in connection with the emission of GW, \citet{Capozziello:2008dv} showed that  the gravitomagnetic corrections on orbital motions  may be significant   in tight binary systems, such as neutron stars or black holes, thus producing peculiar signatures in the GW emission process (see e.g.  \citet{Capozziello:2009eu}).  In addition, gravitomagnetic corrections can be an important tool to constrain the physical properties of astrophysical bodies and investigating the spacetime metric, since they impact on several observables, such as the orbital periods \cite{Iorio:2014yga} or the light bending angle \cite{Ono:2017pie}.}

\section*{Data availability}

Data sharing not applicable to this article as no datasets were generated or analysed during the current study.

\begin{acknowledgments}
The author thanks Dr. Antonello Ortolan for the stimulating and useful discussions.
\end{acknowledgments}

\bibliography{GEM_GW}



\end{document}